# ZnO/Ni interface-activated enhancement of spintronic terahertz emission from Ni/Pt bilayers

Jia Xu[1†*], Shaohua Zhang[1†], Xianguo Jiang[1], Lei Hao[1], Yaxuan Jin[1], Ning Yang[1], Yao Wang[1], Tong Fu[1], Hao Meng[1], Chao Lu[1], Wendeng Huang[1*], Chao Zhou[1*]

[1] Department of Physics, School of Physics and Telecommunication Engineering, Shaanxi University of Technology, Hanzhong 723001, China

[†]These authors contributed equally to this work.

## Abstract

Spintronic terahertz (THz) emitters based on ferromagnet/heavy-metal bilayers provide an efficient route for generating broadband THz radiation through ultrafast spin-to-charge conversion. However, Ni/Pt bilayers usually exhibit relatively weak THz emission, limiting their use as efficient spintronic emitters. Here, we report an interface-assisted enhancement of THz emission in ZnO/Ni/Pt heterostructures. By placing ZnO underneath Ni while preserving the direct Ni/Pt contact, the THz amplitude is increased by up to approximately 2.8 times compared with the Ni/Pt reference. The enhanced emission reverses polarity with the magnetic field and follows the expected angular dependence of ultrafast spin-to-charge conversion. ZnO-, Ni-, and Pt-thickness-dependent measurements show that the emission remains governed by the Ni/Pt conversion process but is strongly affected by the ZnO/Ni bottom interface. Control samples with $Al_2O_3$, Cu, and Fe further exclude a simple underlayer, optical, or growth-quality origin. The enhancement is attributed to a short-range ZnO/Ni interface effect, possibly involving oxygen-related defects, interfacial electric fields, and modification of the Ni interfacial state. This work provides a semiconductor/ferromagnet interface-engineering strategy for activating weak Ni/Pt spintronic THz emitters.

Spintronic terahertz (THz) emitters based on ferromagnetic (FM) / heavy-metal (HM) heterostructures have attracted considerable attention as compact, broadband, and integration-compatible THz sources [1-5]. In a typical spintronic THz emitter, femtosecond laser excitation generates a nonequilibrium spin current in the FM layer, which is injected into the adjacent HM layer and converted into a transverse ultrafast charge current through spin-to-charge conversion (SCC) or orbit-to-charge conversion (LCC) through the inverse spin Hall effect (ISHE) or inverse orbit Hall effect (IOHE) [5-8]. The resulting sub-picosecond charge-current burst radiates a broadband THz pulse. Compared with conventional THz sources based on photoconductive antennas or nonlinear crystals, spintronic THz emitters offer advantages including ultrathin-film geometry, broad spectral bandwidth, magnetic tunability, and compatibility with different pump wavelengths [3, 5, 6, 9].

The efficiency of spintronic THz emission is strongly governed by material selection, layer thickness, stacking order, and interface quality [10-12]. Pt is widely used as an efficient SCC layer because of its strong spin-orbit coupling. However, Ni/Pt bilayers usually exhibit weaker THz emission than optimized Fe/Pt, CoFeB/Pt, or multilayer-based emitters [5, 13, 14]. This relatively weak emission makes Ni/Pt less favorable as a high-output emitter, but it also provides a sensitive platform for exploring additional interface-induced enhancement mechanisms. Moreover, recent studies have shown that Ni-based heterostructures can host rich ultrafast spin and orbital angular-momentum dynamics, suggesting that Ni may be particularly sensitive to interfacial electronic, magnetic, and spin-orbit modifications [15-17].

Interface engineering has become an important strategy for improving spintronic THz emission. Previous studies have mainly focused on optimizing the FM/HM interface, introducing alloyed or impurity-modified heavy metals, designing multilayer structures, or enhancing SCC through interfacial scattering and Rashba-type effects [18-21]. However, the bottom interface of the FM layer, especially a semiconductor (SC) or oxide/FM interface, remains less explored. Such an interface may modify the growth, magnetic boundary condition, interfacial electric field, defect state, and hot-carrier scattering of the FM layer without interrupting the direct FM/HM transport channel [22-24]. Recent studies in $VO_2$/Ni/Pt have shown that integrating a functional oxide layer can strongly modulate THz emission, indicating the potential of oxide/FM interfaces for controlling ultrafast spintronic emission [22].

Among possible SC underlayers, ZnO is a wide-bandgap polar SC with oxygen-related defects, tunable interface states, and strong sensitivity to growth conditions [25, 26]. Although ZnO has been widely studied in optoelectronic and oxide-SC devices, its role in spintronic THz emission, especially as a bottom interface layer beneath Ni, remains largely unexplored. Placing ZnO underneath Ni may therefore provide a SC/FM interface while preserving the direct Ni/Pt contact required for efficient angular-momentum transport and SCC.

In this work, we demonstrate that placing ZnO under Ni can effectively enhance THz emission from Ni/Pt bilayers by up to approximately 2.8-fold. We systematically examine the ZnO-, Ni-, and Pt-thickness dependences, together with magnetic-field, angular, underlayer, and FM-layer control experiments. These measurements show that the enhanced emission remains governed by the Ni/Pt spintronic conversion process, but is strongly modified by the ZnO/Ni bottom interface. The comparison with $Al_2O_3$, Cu, and Fe control samples indicates that the enhancement cannot be attributed solely to improved growth quality or simple optical effects. Instead, it is closely associated with the interfacial properties of ZnO, including oxygen-related defects, interfacial electric fields, and oxide-induced modification of the Ni interfacial state. Our results suggest a SC/FM interface-engineering route for activating weak Ni/Pt spintronic THz emitters.

Bilayer and trilayer samples were deposited on 1-mm-thick glass substrates by magnetron sputtering at room temperature. The layer sequence is described from the substrate side to the top surface. THz emission spectroscopy measurements were performed at room temperature in a dry air environment with relative humidity below 5%. A schematic diagram of the THz emission measurement setup is presented in Fig. 1(a). During measurements, samples were mounted in an electromagnet providing a 1000 Oe in-plane magnetic field along the y-axis. Femtosecond laser pulses were incident normal to the sample surface from the substrate side. Additional details of sample preparation, optical setup, and data analysis are provided in Section S1 of Supplemental Materials.

Figure 1(a) schematically illustrates the THz emission geometry of the ZnO/Ni/Pt heterostructure. In the reference Ni/Pt bilayer, femtosecond laser excitation generates an ultrafast spin current in Ni, which is injected into Pt and converted into an in-plane charge current through ISHE or IOHE [17, 27]. When ZnO is introduced underneath Ni, the ZnO/Ni interface modifies the bottom boundary condition of Ni. This interface may affect Ni growth, interfacial magnetic order, hot-electron scattering, interfacial electric fields, and spin-orbit coupling, thereby increasing the effective ultrafast angular-momentum current reaching Pt [22, 24, 25].

Figure 1(b) compares the time-domain THz waveforms from ZnO(40 nm), Ni(3 nm)/Pt(3 nm) and ZnO(40 nm)/Ni(3 nm)/Pt(3 nm) measured under identical excitation conditions. The ZnO film alone gives a negligible signal, excluding ZnO as an independent THz source in the present geometry. The Ni/Pt bilayer produces a clear but relatively weak THz signal. In contrast, the ZnO/Ni/Pt trilayer shows a much stronger THz response, with amplitude approximately 2.5 times larger than that of the Ni/Pt bilayer. The corresponding frequency-domain spectra, obtained by fast Fourier transform (FFT) of the time-domain signals, are shown in Fig. 1(c). Both Ni/Pt and ZnO/Ni/Pt maintain broadband emission, and the spectral response is also modified after introducing the ZnO underlayer, with a tendency toward broadening of the emission bandwidth. This result indicates that the ZnO layer not only enhances the emission amplitude but also provides an effective route to tune the spectral characteristics of Ni/Pt-based spintronic THz emitters. In addition, ZnO/Ni/Pt was also grown on mica substrates, and the results show that the spectral bandwidth can be modified by changing the substrate, and details are provided in Fig. S2 of Supplemental Materials.

We further measured the THz waveforms under opposite magnetic fields and extracted the peak-to-peak amplitude $\Delta V$, as defined in Fig. 1(b). The extracted $\Delta V$ values are summarized in Fig. 1(d) (See details in Fig. S3 of Supplemental Materials). Reversing the magnetic field reverses the THz polarity for both Ni/Pt and ZnO/Ni/Pt while maintaining nearly symmetric amplitudes. This behavior confirms that the THz emission is magnetization-dependent. Figure 1(e) compares the waveforms under front-side and back-side excitation. The corresponding response remains consistent with spintronic emission after considering the reversal of excitation geometry, which helps exclude simple pump-direction-dependent optical artifacts. The magnetic-field-dependent THz amplitude in Fig. 1(f) follows the typical magnetic switching behavior and saturates well below the magnetic field of 1000 Oe used in the measurements. These results demonstrate that the ZnO/Ni/Pt trilayer remains a spintronic THz emitter, but with an enhanced emission responses compared with Ni/Pt.

We next examined the angular dependence and ZnO-thickness dependence to clarify the origin of the enhancement. Figure 2(a) shows the sample-azimuthal-angle-dependent THz amplitude, the corresponding THz signal spectra are shown in Fig. S4 of Supplemental Materials. The nearly angle-independent response indicates that the enhancement is not caused by macroscopic in-plane anisotropy of the sample. Figure 2(b) presents the THz peak amplitude as a function of the in-plane magnetic-field angle. Both Ni/Pt and ZnO/Ni/Pt exhibit a sinusoidal angular dependence, consistent with an ultrafast transverse charge current generated by SCC [28-30]. The ZnO/Ni/Pt trilayer retains the same angular symmetry as Ni/Pt but with a larger amplitude. This result indicates that the ZnO underlayer enhances the spintronic emission channel rather than producing a different angular-symmetry current. In addition, we measured Pt/Ni and ZnO/Pt/Ni samples, and similar enhancement can be observed in structures containing ZnO. The corresponding details are shown in Fig. S5 of Supplemental Materials.

We further investigated the ZnO-thickness-dependent THz emission, as shown in Fig. 2(c). The time-domain THz waveforms of ZnO($d_{\mathrm{ZnO}}$)/Ni(3 nm)/Pt(3 nm) show that the THz signal increases rapidly once the ZnO layer is introduced. To quantify this behavior, the peak-to-peak amplitude $\Delta V$ was extracted and plotted as a function of ZnO thickness in Fig. 2(d). Here, the error bars in Fig. 2 (d) represent the standard errors of peak-to-peak THz amplitudes obtained from time-domain signals after baseline subtraction and Gaussian fitting. Further increasing the ZnO thickness produces only weak changes in the amplitude, and the signal tends to saturate. This rapid onset and weak ZnO-thickness dependence in thicker films indicate that the enhancement is governed by the ZnO/Ni interface rather than by the bulk ZnO layer. If the enhancement were dominated by optical absorption,photocarriers, or THz out-coupling in ZnO, a more continuous dependence on ZnO thickness would be expected. Therefore, Fig. 2 identifies the ZnO/Ni interface as the key ingredient for the enhanced emission.

To further understand how the ZnO/Ni interface modifies the Ni/Pt emitter, we measured the Ni- and Pt-thickness-dependent THz emission. Figure 3(a) shows the time-domain THz waveforms from ZnO(40 nm)/Ni($d_{\mathrm{Ni}}$)/Pt(3 nm) and Ni($d_{\mathrm{Ni}}$)/Pt(3 nm). The corresponding peak-to-peak amplitudes $\Delta V$ are summarized in Fig. 3(b). Both series exhibit a nonmonotonic dependence on Ni thickness, reflecting the competition among optical absorption, ultrafast angular-momentum generation, spin relaxation, and transport across the Ni/Pt interface [15, 31]. Compared with the Ni/Pt reference, the ZnO/Ni/Pt trilayers show larger THz amplitudes over most of the measured Ni-thickness range. This enhancement is more pronounced in the thin-Ni regime and gradually becomes weaker as the Ni layer becomes thicker, indicating that the ZnO underlayer mainly modifies the interfacial or near-interface region of Ni.

The Pt-thickness-dependent THz emission was also examined, as shown in Figs. 3(c) and 3(d). Both ZnO(40 nm)/Ni(3 nm)/Pt($d_{\mathrm{Pt}}$) and Ni(3 nm)/Pt($d_{\mathrm{Pt}}$) exhibit a nonmonotonic dependence on Pt thickness. The THz amplitude first increases with increasing Pt thickness and then decreases at larger thicknesses. This behavior is consistent with the role of Pt as the main SCC layer, where the emitted THz amplitude is governed by the balance between SCC, spin relaxation, current shunting, and THz attenuation [10, 13, 32]. In the optimized Ni- and Pt-thickness regions, the ZnO/Ni/Pt series shows a THz response up to approximately 2.8 times larger than that of Ni/Pt, indicating that the ZnO/Ni interface enhances the effective ultrafast angular-momentum current injected into Pt. The persistence of both Ni- and Pt-thickness dependences further indicates that the enhanced emission is not generated by ZnO alone, but remains coupled to the Ni/Pt spintronic conversion process.

To clarify whether the enhancement is specific to the ZnO/Ni interface, we performed a series of control experiments, as summarized in Fig. 4. Figure 4(a) compares the THz emission from Ni(3 nm)/Pt(3 nm) with that from Ni/Pt samples grown on different 2-nm-thick underlayers, including ZnO, $Al_2O_3$, and Cu. The ZnO underlayer produces the strongest enhancement. $Al_2O_3$ underlayer also moderately enhances the Ni/Pt emission, which may be partly related to improved growth quality of the ultrathin Ni/Pt film on an oxide seed layer. In contrast, the Cu underlayer does not lead to a comparable enhancement and even slightly reduces the emission, possibly due to additional optical absorption or current shunting in the metallic underlayer [10, 25, 33].

We further measured the X-ray reflectivity rocking curves of Ni(3 nm)/Pt(3 nm) and ZnO(40 nm)/Ni(3 nm)/Pt(3 nm), as shown in Fig. S6 of the Supplemental Materials. The comparable peak widths suggest that the ZnO-induced enhancement cannot be simply attributed to a global improvement of film crystallinity or growth quality. This comparison indicates that the enhanced emission is not caused merely by introducing an additional underlayer. Instead, an oxide/Ni interface is beneficial for activating the Ni/Pt emitter, while a metallic Cu/Ni bottom interface cannot reproduce this effect. The stronger enhancement in ZnO/Ni/Pt may be related to the polar nature of ZnO, interface-induced electric fields, oxygen-related defects, or a stronger modification of the interfacial electronic and magnetic states of Ni.

This interpretation is further supported by the ZnO growth-condition dependence and annealing experiments shown in the Figs. S7-S9 of Supplemental Materials. The THz amplitude increases when a higher oxygen fraction is introduced during ZnO growth, whereas under pure Ar growth the THz amplitude decreases with increasing Ar pressure. In addition, the THz signal of ZnO/Ni/Pt is markedly reduced after annealing at 300 °C for 20 min. These trends indicate that oxygen-related defects, film density, and interface quality of ZnO are closely associated with the enhanced THz emission.

The short-range nature of the ZnO/Ni interface effect was further verified by inserting a Cu spacer between ZnO and Ni. As shown in Fig. 4(b), the THz amplitude of ZnO/Cu/Ni/Pt is strongly reduced compared with that of ZnO/Ni/Pt and becomes close to the Ni/Pt reference (See details in Fig. S10 of Supplemental Materials). This result demonstrates that direct ZnO/Ni interface is required for the enhancement. Therefore, the enhanced emission cannot be explained by a substrate effect, optical absorption of ZnO, or THz optical effects. Instead, it originates from a local interfacial modification of Ni by the underlying ZnO layer.

We also examined whether a similar enhancement can be obtained in Fe-based structures. Figure 4(c) compares Fe(3 nm)/Pt(3 nm), ZnO(40 nm)/Fe(3 nm), and ZnO(40 nm)/Fe(3 nm) /Pt(3 nm). The ZnO/Fe bilayer produces a negligible THz signal, excluding the SCC or LCC at the ZnO/Fe interface as a source for the enhancement. ZnO/Fe/Pt does not exhibit a comparable enhancement relative to Fe/Pt, as confirmed by the other measurements shown in Fig. S11 of Supplemental Materials. This behavior is quite different from the strong enhancement observed in ZnO/Ni/Pt, indicating that the oxide-interface-induced enhancement is most effective for Ni/Pt. The Ni selectivity suggests that the ZnO/Ni interface couples to the specific electronic and magnetic properties of Ni. In particular, Ni may be more sensitive to oxide-interface-induced modifications of magnetic order, hot-carrier transport, and spin-orbit-coupled angular-momentum dynamics than Fe under the present experimental conditions. In addition, the absence of enhancement when Ni was replaced by Co also highlights the uniqueness of Ni/ZnO interface, instead of a generic effect existent for all ZnO/FM systems (See details in Fig. S12 of Supplemental Materials).

Based on these results, the proposed physical picture is illustrated in Fig. 4(d). In the reference Ni/Pt bilayer, femtosecond laser excitation generates an ultrafast angular-momentum current in Ni, which is injected into Pt and converted into a transverse charge current through SCC, giving rise to THz radiation. When ZnO is introduced underneath Ni, the ZnO/Ni interface modifies the bottom boundary condition of the Ni layer. Broken inversion symmetry, interfacial electric fields, oxygen-related defects, and oxide-induced changes in Ni growth may enhance the generation or injection of ultrafast spin and orbital angular-momentum currents [23, 34-36]. These enhanced angular-momentum currents are then transported into Pt, where they are converted into a larger transient charge current through the ISHE and possibly orbital-related conversion channels. The resulting stronger transient charge current leads to the enhanced THz emission from ZnO/Ni/Pt. Therefore, the ZnO underlayer acts as an interface-activation layer for the weak Ni/Pt spintronic THz emitter.

In conclusion, we have demonstrated that introducing a ZnO underlayer effectively enhances the spintronic THz emission from Ni/Pt bilayers by up to approximately 2.8-fold. The enhanced emission retains the magnetic-field and angular dependences expected for ultrafast SCC, confirming its spintronic origin. Thickness-dependent measurements show that the ZnO/Ni/Pt emission remains governed by the Ni/Pt spintronic conversion process, while the ZnO/Ni interface modifies the effective ultrafast angular-momentum generation and transport. Control experiments using $Al_2O_3$, Cu, and Fe further reveal that the enhancement is associated with a short-range oxide/Ni interface effect rather than bulk ZnO emission, simple optical absorption, or a general underlayer effect. Our results identify the ZnO/Ni interface as an activation layer for weak Ni/Pt THz emitters and provide a practical SC/magnetic-heterostructure interface-engineering strategy for improving Ni-based spintronic THz sources.

See the supplemental materials for details on sample fabrication, magnetic-field-direction dependence of the THz spectra, stacking-sequence and thickness dependence, the influence of FM layers on THz emission in trilayer structures, and the measurement of the X-ray reflectivity curves of the samples.

This work was supported by the National Natural Science Foundation of China (Grant No. 12204295 and 12204296), the Open Research Project of State Key Laboratory of Surface Physics from Fudan University (Grant No. KF2022_17), the Youth Science and Technology Star Program of Shaanxi Province (2025ZC-KJXX-15), the Young Talent Fund of Association for Science and Technology in Shaanxi (20250512), the Youth Project in Sanqin Talent Introduction Program, and the Shaanxi Province Outstanding Young Talent Support Program for Universities and Young Hanjiang Scholar of Shaanxi University of Technology, and Shaanxi University of Technology (Grants No. SLGRCQD044 and SLGRCQD046). The Innovation Fund of Shaanxi University of Technology (Grant No. SLGYCX2631, SLGYCX2511 and SLGYCX2514). Hao Meng acknowledges the support from the National Natural Science Foundation of China (Grant No.12174238).

## AUTHOR DECLARATIONS

### Conflict of Interest

The authors have no conflicts to disclose.

### Author Contributions

**Jia Xu**: Conceptualization, methodology, investigation, writing - original draft, funding acquisition. **Shaohua Zhang**: Data curation, investigation, writing - original draft. **Xianguo Jiang**: Data curation, investigation. **Lei Hao**: Methodology, formal analysis. **Yaxuan Jin**: Methodology, formal analysis. **Ning Yang**: Methodology. **Yao Wang:** Review and editing. **Tong Fu**: Review and editing. **Hao Meng**: Review and editing. **Chao Lu**: Review and editing. **Wendeng Huang**: Review and editing. **Chao Zhou**: Conceptualization, methodology, investigation, writing - original draft, funding acquisition.

## DATA AVAILABILITY

The data that support the findings of this study are available from the corresponding authors upon reasonable request.

**Figures:**

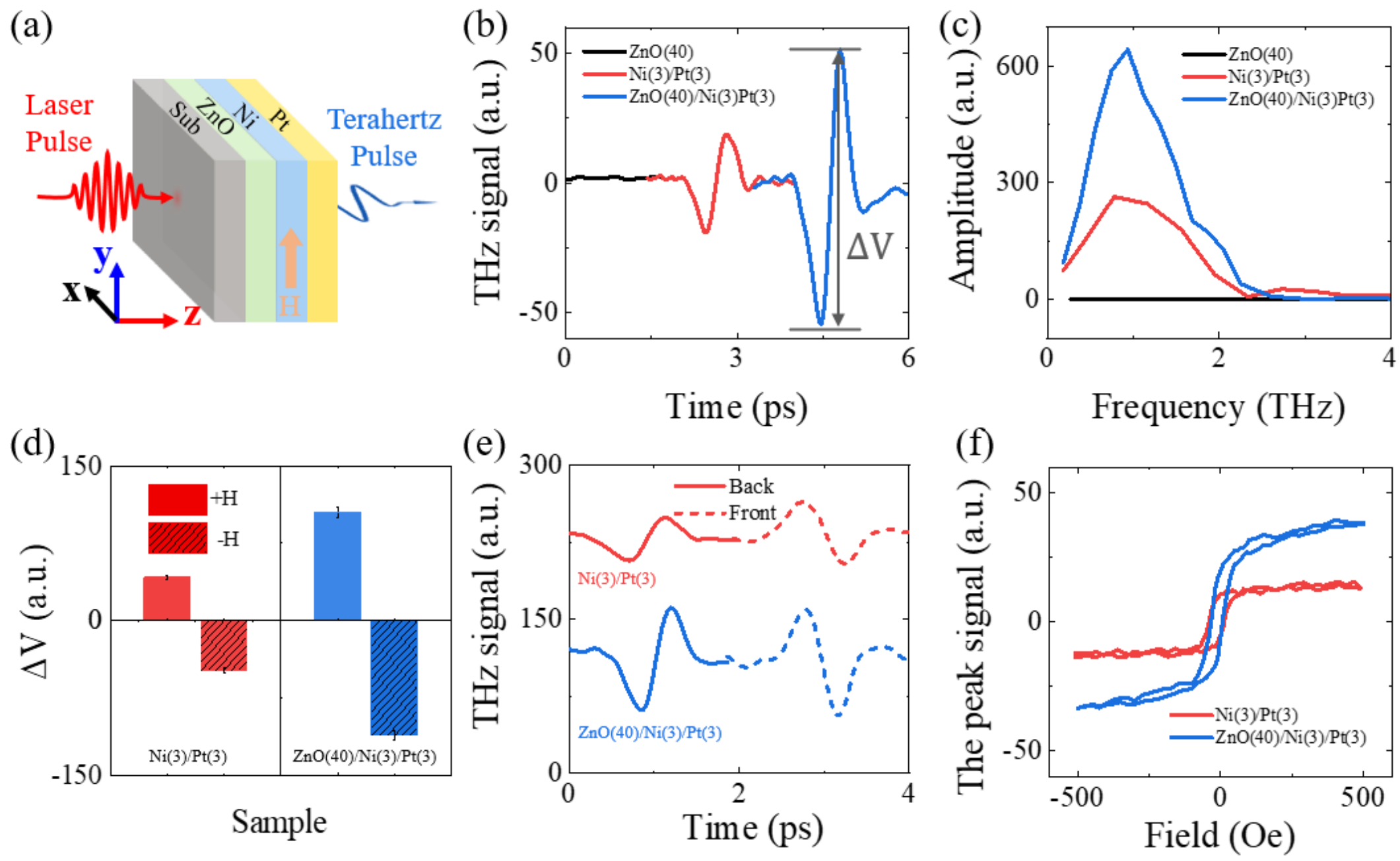

**Figure 1** (a) Schematic of the ZnO/Ni/Pt structure and THz emission geometry under substrate-side femtosecond excitation. (b) Time-domain THz waveforms from ZnO(40 nm), Ni(3 nm)/Pt(3 nm), and ZnO(40 nm)/Ni(3 nm)/Pt(3 nm) measured under identical conditions. (c) Corresponding frequency-domain spectra for time-domain signals in (b). (d) Extracted peak-to-peak amplitudes under opposite magnetic fields, with $\Delta V$ defined in (b). (e) THz waveforms under front-side and back-side excitation for Ni(3 nm)/Pt(3 nm) and ZnO(40 nm)/Ni(3 nm)/Pt(3 nm) under in-plane field. (f) Magnetic-field dependence of the THz peak signal.

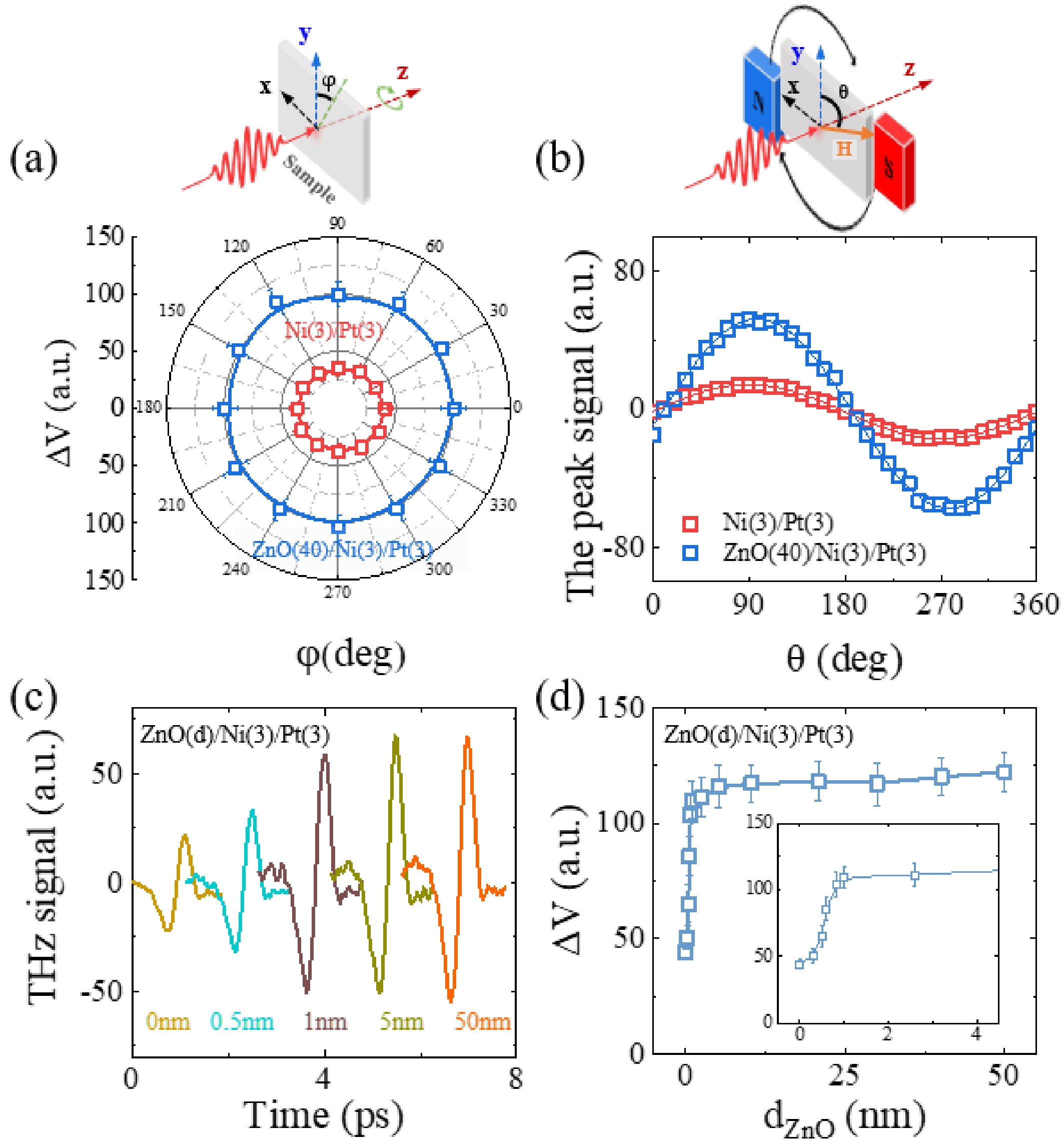

**Figure 2** (a) Sample-azimuthal-angle dependence of $\Delta V$ for Ni(3 nm)/Pt(3 nm) and ZnO(40 nm)/Ni(3 nm)/Pt(3 nm). (b) Magnetic-field-angle dependence of the THz peak signal. (c) Time-domain THz waveforms from ZnO($d$)/Ni(3 nm)/Pt(3 nm). The curves have been shifted horizontally for clarity. (d) Extracted $\Delta V$ as a function of ZnO thickness. The inset shows the thin-thickness region from 0 to 4 nm. The solid lines are guides to the eye.

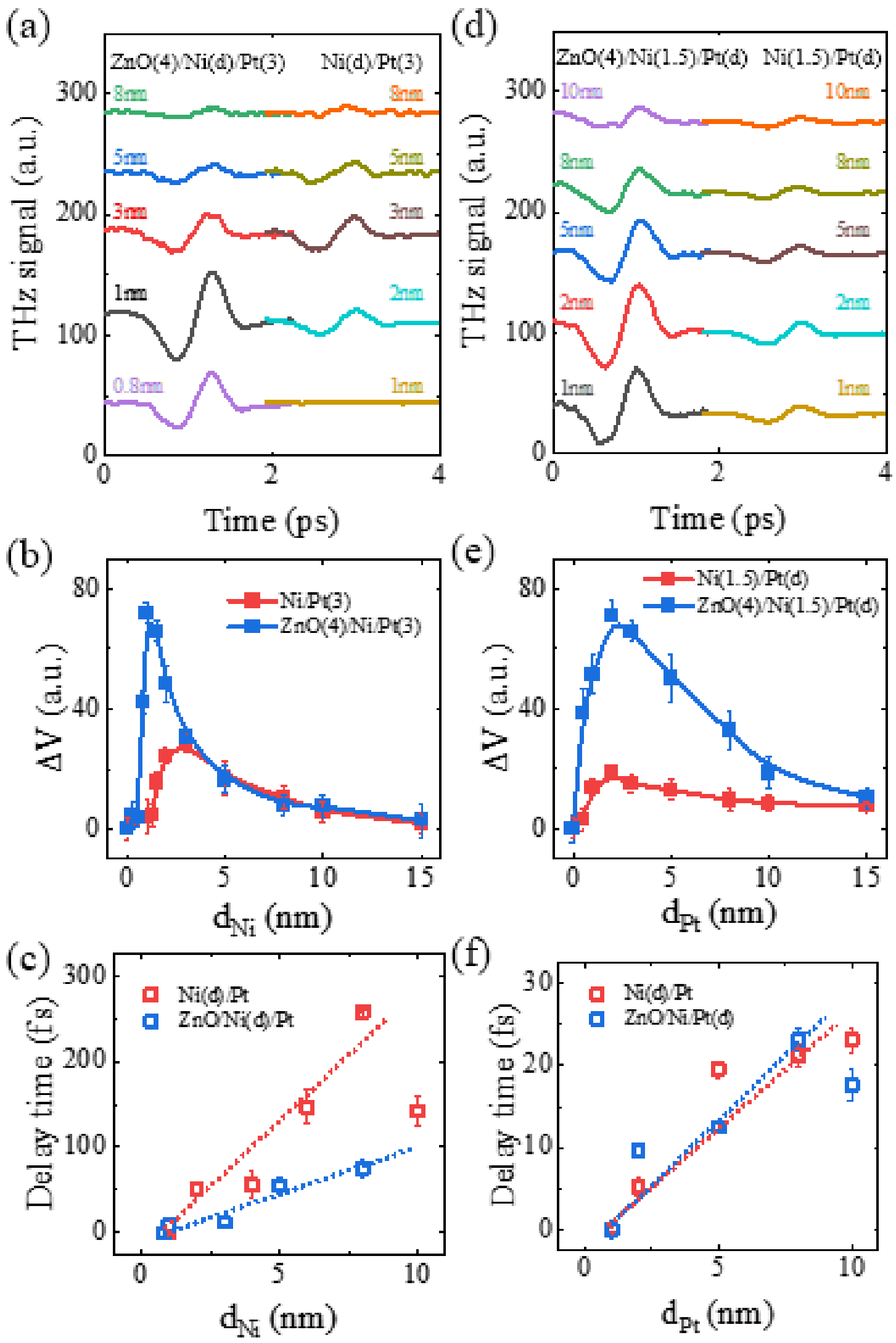

**Figure 3** (a) Time-domain THz waveforms from ZnO(40 nm)/Ni(*d*)/Pt(3 nm) and Ni(*d*)/Pt(3 nm) with varying Ni thickness. (b) Extracted $\Delta V$ as a function of Ni thickness for corresponding bilayer and trilayer. (c) Delay time $\tau_D$ extracted by Hilbert transform for bilayer samples as a function of Ni thickness of ZnO(40 nm)/Ni(*d*)/Pt(3 nm) and Ni(*d*)/Pt(3 nm). (d)Time-domain THz waveforms from ZnO(40 nm)/Ni(3 nm)/Pt(*d*) and Ni(3 nm)/Pt(*d*) with varying Pt thickness. (e) Extracted $\Delta V$ as a function of Pt thickness for corresponding bilayer and trilayer. The solid lines in (b) and (e) are guides to the eye. (f) Delay time $\tau_D$ extracted by Hilbert transform for bilayer samples as a function of Pt thickness of ZnO(40 nm)/Ni(3 nm)/Pt(*d*) and Ni(3 nm)/Pt(*d*).

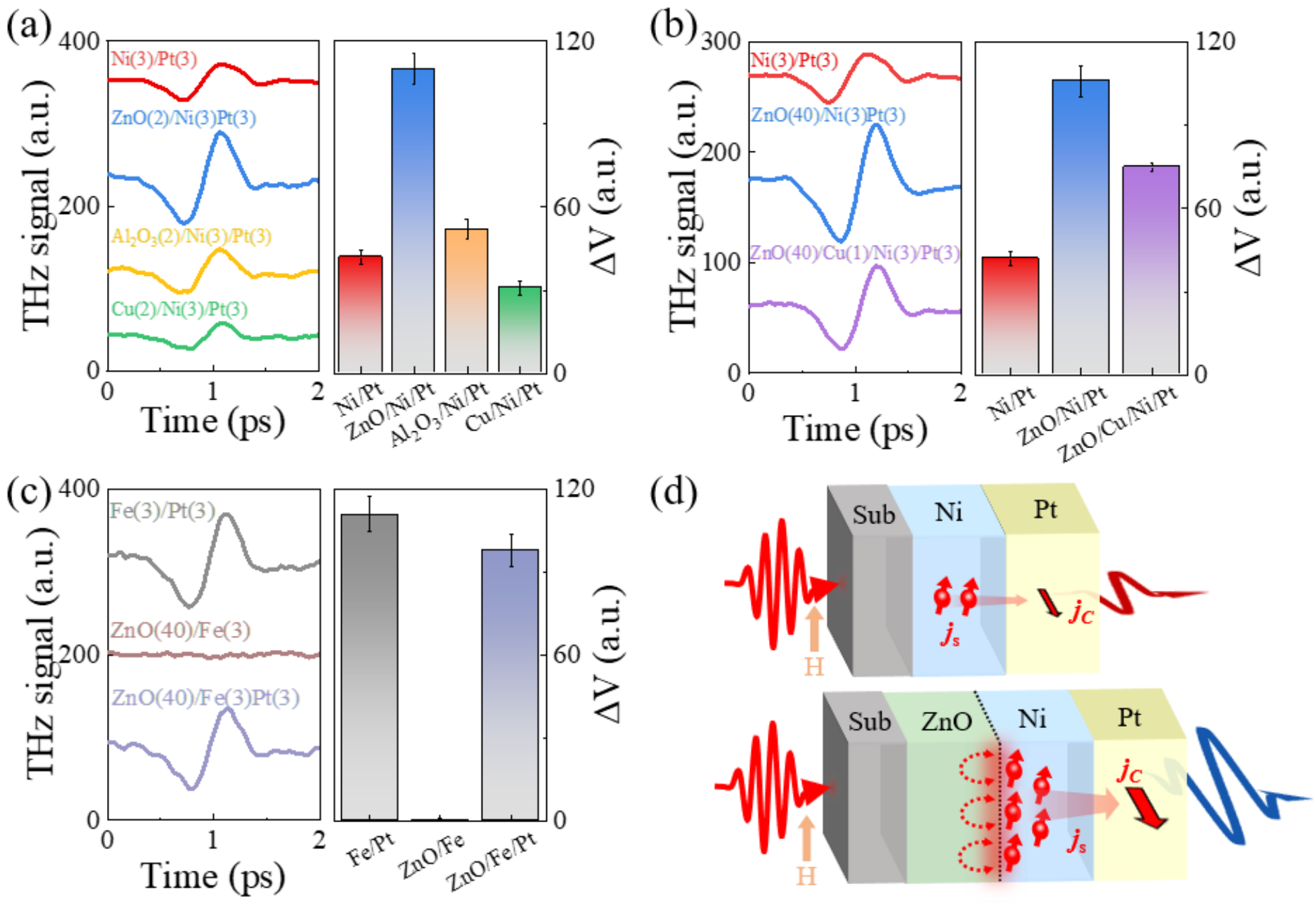

**Figure 4** (a)-(c) THz waveforms and extracted $\Delta V$ for (a) Ni(3 nm)/Pt(3 nm), ZnO(2 nm)/Ni(3 nm)/Pt(3 nm), $Al_2O_3$(2 nm)/ Ni(3 nm)/Pt(3 nm), and Cu(2 nm)/ Ni(3 nm)/Pt(3 nm), (b) Ni(3 nm)/Pt(3 nm), ZnO(40 nm)/Ni(3 nm)/Pt(3 nm), and ZnO(40 nm)/Cu(1 nm)/Ni(3 nm)/Pt(3 nm). (c) Fe(3 nm)/Pt(3 nm), ZnO(40 nm)/Fe(3 nm), and ZnO(40 nm)/Fe(3 nm)/Pt(3 nm). (d) Schematic of the proposed mechanism of ZnO/Ni-interface-assisted enhancement.